\documentclass[twocolumn,epjc3]{svjour3}
 \usepackage{graphicx,amsmath,amssymb}
 \usepackage{units}
 \usepackage{axodraw}

 \newcommand{\jouref}[4]{{#1 }{\bf #2} (#3) #4}

 \journalname{Eur. Phys. J. C}

\begin{document}

 \title{Soft confinement in a 3-d spin system}

 \author{K.~Odagiri \and T.~Yanagisawa}

 \institute{
  Electronics and Photonics Research Institute,
  National Institute of Advanced Industrial Science and Technology,
  Tsukuba Central 2,
  1--1--1 Umezono, Tsukuba, Ibaraki 305--8568, Japan
  }

 \date{\today}

 \maketitle

 \begin{abstract}
  We consider a 1+3 dimensional spin system.
  The spin-wave (magnon) field is described by the O(3) non-linear sigma 
model with a symmetry-breaking potential.
  This interacts with a slow spin SU(2) doublet Schr\"{o}dinger fermion.
  The interaction is described by a generalized nonperturbative Yukawa 
coupling, and the self-consistency condition is solved with the aid of a 
non-relativistic Gribov equation.
  When the Yukawa coupling is sufficiently strong, the solution exhibits 
supercriticality and soft confinement, in a way that is quite analogous 
to Gribov's light-quark confinement theory.

  The solution corresponds to a new type of spin polaron, whose 
condensation may lead to exotic superconductivity.

 \end{abstract}


 \section{Introduction}

 \subsection{Motivation}

  The motivation for our work is twofold.

  First, we have the colour confinement problem, to which one suggested 
solution is Gribov's light-quark confinement theory 
\cite{gribovlargeshort,gribovquarkconfinement,gribovlectures,yurireview}.
  The discovery of an analogy or application for it in another system 
will be beneficial for deepening our understanding about light-quark 
confinement theory and confinement in general.

  Second, the behaviour of fermionic spin in a polarized background is 
often described by spin polarons \cite{spinpolaron}, which are known to 
occur in some limiting strongly correlated cases. It would be 
interesting to find some other instances of spin physics in which 
fermionic spin is strongly affected by the polarized background.

 \subsection{General remarks}

  When discussing the colour confinement problem, one traditional 
approach is to look primarily at the gluodynamics at large scales.
  As a representative example, we have Wilson's work 
\cite{wilsonconfinement} which is based on lattice quantization.
  In this picture, in simplest terms, coloured objects are confined 
because the long-distance effective potential is unbounded from above.
  When a quark--antiquark pair is pulled apart, for example, it requires 
infinite energy to separate them by an infinite distance.

  However, in the real world, there are light quarks.
  When two quarks are pulled apart, they will fragment into hadrons.
  This being the case, we arrive at an alternative line of thought, that 
large long-distance interaction is not necessarily the true essence of 
confinement as it relates to our world.
  What would then be the condition that governs confinement?

  This problem is dealt with by Gribov's light-quark confinement theory 
\cite{gribovlargeshort,gribovquarkconfinement,gribovlectures,yurireview}, 
according to which confinement occurs when moderately strong coupling 
binds together light quarks supercritically.
  The supercriticality changes the vacuum structure, and the response of 
the modified vacuum is described by the contribution of Goldstone pions. 

  A somewhat unobvious outcome of this picture is that although 
moderately strong gluonic interaction is responsible for changing the 
vacuum structure, it is the interaction of pions that confines the 
quarks.
  One intuitive explanation for this is that when a quark--antiquark 
pair is pulled apart, they fragment into more mesons such as pions, with 
no reference to dynamics that grows large at long distances.
  Hadrons are fragile objects.

  Let us call the first picture `hard confinement' and the second 
picture `soft confinement' \cite{yurireview} (note, however, that the 
first picture is called `soft' in ref.~\cite{wilsonconfinement}).
  We would now like to look for their counterparts in spin systems.

  An analogy for hard confinement is in the so-called `string polarons' 
\cite{stringpolaron}, which occurs in two-dimensional antiferromagnetic 
systems with small spin exchange energy $J$ at low carrier (electron or 
hole) concentration.
  Here, moving a single carrier over the antiferromagnetic background 
disturbs the spin.
  The amount of disturbance is proportional to the distance $\ell$ moved 
by the carrier, and therefore there is an effective potential 
$V(\ell)\propto\ell$ which traps the carrier, or confines it into 
spin-singlet pairs.

  Our question is whether there is any such analogy for soft 
confinement.
  Pions being Goldstone bosons, and magnons which are the quanta of spin 
wave being Goldstone bosons, we are led to consider magnon-fermion 
systems.
  The most obvious candidate is then the three-dimensional spin-wave 
system with linear dispersion relation, which is coupled to a 
spin-doublet carrier through a Yukawa-like interaction.

  We shall study this system, and shall show that soft confinement does 
indeed occur when the interaction is strong.
  This solution corresponds to a new type of spin polaron state, in 
which fermionic spin is confined not by large spatial long-distance 
interaction but by the temporal decay of the false vacuum.
  More phenomenologically speaking, a fermionic spin that is excited 
will decay by emitting a magnon, and will not have time to behave like 
an itinerant spin.

  Our work utilizes Gribov equations 
\cite{gribovlargeshort,gribovquarkconfinement,gribovlectures,yurireview}, 
but they are used here for somewhat different set of reasons than in the 
QCD case.

  We shall formulate the problem in Sec.~\ref{sec_formulation}.
  We shall write down the Gribov equations in Sec.~\ref{sec_griboveqn} 
and solve them in Sec.~\ref{sec_gribov_solution}.
  We analyze the solutions corresponding to both weak and strong 
coupling regions in Sec.~\ref{sec_analysis}.
  The nature of the supercritical solutions is discussed in 
Sec.~\ref{sec_discussion}, together with some phenomenological remarks 
and a comparison with QCD.
  The conclusions are stated at the end.

 \section{Formulation of the problem}
 \label{sec_formulation}

  The spin-wave system is described by the following Lagrangian density:
 \begin{equation}
  \mathcal{L}_\Phi=\frac{\hbar^2}2\left[
   \left(\frac{\partial\vec\Phi}{\partial t}\right)^2-
   u^2\left(\nabla\vec\Phi\right)^2
  \right]-V(\vec\Phi).
  \label{eqn_nlsm_lagrangian}
 \end{equation}
  $u$, which adopts the role of $c$ or the speed of light, is the 
spin-wave velocity.
  $V(\vec\Phi)$ is a potential such as 
 \(
-\nicefrac\mu2\left|\vec\Phi\right|^2+\nicefrac\lambda4\left|\vec\Phi\right|^4, 
 \)
  whose minimum is found at $\left|\vec\Phi\right|=v_h$.
  $v_h$ is non-zero, and this implies finite magnetization.
  $\vec\Phi$ may then be parametrized as
 \begin{equation}
  \vec\Phi=(\phi_1,\phi_2,v_h+h).
  \label{eqn_phi_parametrization}
 \end{equation}
  The magnon Green's function may then be written as
 \begin{equation}
  D_\pm(\omega,\mathbf{k})=\frac1{\omega^2-u^2\mathbf{k}^2+i0}.
 \end{equation}
  $\pm$ refers to the $S_z=\pm1$ magnon modes, i.e.\ $\phi_x\pm i\phi_y$ 
when magnetization is along the $z$ axis.
  In principle, there will also be the amplitude mode which we may 
denote
 \begin{equation}
  D_0(\omega,\mathbf{k})=
  \frac1{\omega^2-u^2\mathbf{k}^2-M_h^2u^4+i0},
 \end{equation}
  but we shall not consider the contributions of this mode in this 
study.
  That is, $M_hu^2$ will be assumed to be large.

  The fermion $\Psi$ is assumed to move at a velocity $\sim 
v_\mathrm{F}$ that is much smaller than $u$, so that the 
momentum-dependent terms may be neglected.
  The Lagrangian is given by
 \begin{equation}
  \mathcal{L}_\Psi= \Psi^\dagger\left[
  T\left(i\hbar\frac{d}{dt}\right)
  -\mu_\mathrm{F}+f^{-1}\Delta_\mathrm{ex}
  \vec\sigma\cdot\vec\Phi\right]\Psi.
  \label{eqn_yukawa_lagrangian}
 \end{equation}
  $T$ and $\Delta_\mathrm{ex}$ (generalized exchange energy) are some 
functions of the time derivative.
  $\Psi$ is an SU(2) doublet.
  $f=2v_h$ is the form factor of the magnon, and has the dimension 
$[$energy$\times$volume$]^{-1/2}$.
  $\mu_\mathrm{F}$ is the Fermi energy.
  When $\vec\Phi$ is as parametrized by 
eqn.~(\ref{eqn_phi_parametrization}), the $S_z=\pm\nicefrac12$ spin 
states $\psi_\pm$ are given by
 \begin{equation}
  \Psi=\left(\begin{array}{c}\psi_+\\\psi_-\end{array}\right).
 \end{equation}
  The $\psi_-$ state is more energetically favourable.

  Equation (\ref{eqn_yukawa_lagrangian}) generalizes the weak-coupling 
expression
 \begin{equation}
  \mathcal{L}_\Psi= \Psi^\dagger\left[
  i\hbar\frac{d}{dt}-\mu_\mathrm{F}+f^{-1}\Delta_0
  \vec\sigma\cdot\vec\Phi\right]\Psi.
  \label{eqn_yukawa_lagrangian_weak}
 \end{equation}
  The exchange energy $\Delta_0$ is constant.
  In this case, the Green's function is given by
 \begin{equation}
  G^\mathrm{weak}_\pm(\omega)=
  \frac1{\omega-\mu_\mathrm{F}\mp\Delta_0/2}.
  \label{eqn_g_weak}
 \end{equation}
  We see that $\Delta_0=G_-^{-1}(\omega)-G_+^{-1}(\omega)$.
  We then generalize this result as
 \begin{equation}
  \Delta_\mathrm{ex}(\omega)=G^{-1}_-(\omega)-G^{-1}_+(\omega).
  \label{eqn_delta_ex_general}
 \end{equation}
  Note that in the weak-coupling expression of 
eqn.~(\ref{eqn_yukawa_lagrangian_weak}), the interaction term may be 
rotated away by the transformation
 \begin{equation}
  \Psi(x,t)\longrightarrow\exp\left(
  \int \frac{i}{\hbar}f^{-1}\Delta_0\vec\sigma\cdot\vec\Phi(x,t) dt
  \right)\Psi(x,t).
 \end{equation}
  This is not the case when the interaction is strong.

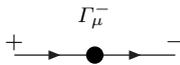
\begin{figure}[ht]
  \centerline{
  \begin{picture}(100,60)(0,0)
   \ArrowLine(20,30)(50,30)
   \ArrowLine(50,30)(80,30)
   \GCirc(50,30){3}{0}
   \Text(20,35)[c]{$+$}
   \Text(80,35)[c]{$-$}
   \Text(50,45)[c]{$\Gamma_\mu^-$}
  \end{picture}
  }
  \caption{\label{fig_current_simple}
  Off-diagonal spin current. There is a spin $S_z$ conservation law at 
the vertex. Fermions carry $\pm\nicefrac12$ $S_z$ charge, whereas the 
magnon has $S_z=\pm1$.}
 \end{figure}

  Our writing eqn.~(\ref{eqn_delta_ex_general}) is based on the 
conservation of spin current.
  Let us consider a current of the form shown in 
fig.~\ref{fig_current_simple}. Note that $S_z$ is conserved.

  Whatever is the form of $\Gamma_\mu$, its contraction with the 
incoming 4-momentum $q_\mu$ will be zero in the soft limit $q_\mu\to0$. 
In this limit, the violation of the Ward identity will be proportional 
to $G^{-1}_--G^{-1}_+$. This must be cancelled by the magnon 
contribution, and therefore the coupling is of the form of 
eqn.~(\ref{eqn_yukawa_lagrangian}) together with 
eqn.~(\ref{eqn_delta_ex_general}).

\begin{figure}[ht]{
  \centerline{
    \begin{picture}(120,60)(0,0)
     \ArrowLine(20,20)(40,20)
     \ArrowLine(40,20)(80,20)
     \ArrowLine(80,20)(100,20)
     \DashCArc(60,20)(20,0,180){5}
     \Text(10,10)[c]{$\psi_+(q)$}
     \Text(55,10)[c]{$\psi_-(k)$}
     \Text(60,50)[c]{$\phi_+(q-k)$}
     \Text(100,10)[c]{$\psi_+(q)$}
    \end{picture}
  }
  \caption{The self-energy diagram for $\psi_+$.
  \label{fig_one_loop_electron_self_energy}}}
 \end{figure}
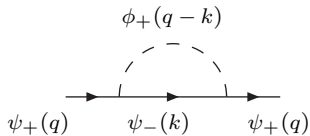

  Let us now consider the one-loop self-energy diagram, shown in 
fig.~\ref{fig_one_loop_electron_self_energy}.

  We adopt the four-vector notation with the metric 
$\mathrm{diag}(1,-1,-1,-1)$ when $u=c=1$.

  At the dressed one-loop level, and corresponding to 
fig.~\ref{fig_one_loop_electron_self_energy}, the self-energy is given 
by
 \begin{equation}
  \Sigma_+(q)=\int\frac{d^4k}{(2\pi)^4i}
  f^{-2}\Delta_\mathrm{ex}^2(k,q) G_-(k)D_\pm(q-k).
  \label{eq:electron_self_energy_dressed}
 \end{equation}
  $\Sigma_+$ refers to the correction to $G^{-1}_+$.
  $\Sigma_-$ is given by
 \begin{equation}
  \Sigma_-(q)=\int\frac{d^4k}{(2\pi)^4i}
  f^{-2}\Delta_\mathrm{ex}^2(k,q) G_+(k)D_\pm(q-k).
  \label{eq:electron_self_energy_dressed_another}
 \end{equation}

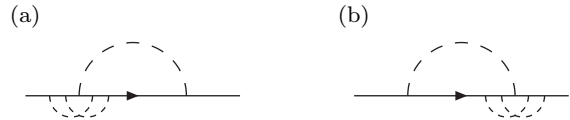
\begin{figure}[ht]{
  \centerline{
    \begin{picture}(120,60)(0,0)
     \Text(20,50)[c]{(a)}
     \ArrowLine(20,20)(100,20)
     \DashCArc(60,20)(20,0,180){5}
     \DashCArc(37,20)(8,180,360){2.5}
     \DashCArc(43,20)(8,180,360){2.5}
    \end{picture}
    \begin{picture}(120,60)(0,0)
     \Text(20,50)[c]{(b)}
     \ArrowLine(20,20)(100,20)
     \DashCArc(60,20)(20,0,180){5}
     \DashCArc(77,20)(8,180,360){2.5}
     \DashCArc(83,20)(8,180,360){2.5}
    \end{picture}
  }
  \caption{The two perturbative three-loop self-energy diagrams a and b 
are equivalent.
  \label{fig_one_loop_electron_self_energy_double}}}
 \end{figure}

  This set of equations needs special emphasis.
  Suppose that, instead of the above formulation, we had started from 
the weak coupling formulation of eqn.~(\ref{eqn_yukawa_lagrangian_weak}) 
and then defined the full theory based on some perturbative expansion.
  In that case, if we define $\Delta_\mathrm{ex}$ perturbatively as the 
bare vertex plus corrections, then starting at the three-loop level, 
there will apparently be double counting in 
eqns.~(\ref{eq:electron_self_energy_dressed},%
\ref{eq:electron_self_energy_dressed_another}), 
as shown in fig.~\ref{fig_one_loop_electron_self_energy_double}.
  One would think that each of 
fig.~\ref{fig_one_loop_electron_self_energy_double}a, b corrects one or 
the other of the vertex and therefore one or the other of 
$\Delta_\mathrm{ex}$ whereas the two diagrams are in fact equivalent.

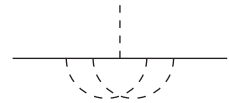
\begin{figure}[ht]{
  \centerline{
    \begin{picture}(120,80)(0,0)
     \Line(20,40)(100,40)
     \DashCArc(55,40)(15,180,360){3}
     \DashCArc(65,40)(15,180,360){3}
     \DashLine(60,40)(60,60){3}
    \end{picture}
  }
  \caption{The leading perturbative vertex correction contribution.
  \label{fig_vertex_correction}}}
 \end{figure}

  However, this is not quite true. Diagrams of the form shown in 
fig.~\ref{fig_one_loop_electron_self_energy_double} cannot be part of 
the renormalization of $\Delta_\mathrm{ex}$, if $\Delta_\mathrm{ex}$ is 
defined by eqn.~(\ref{eqn_delta_ex_general}).
  For instance, the leading self-energy correction contribution starts 
at the one-loop order as can be seen in 
fig.~\ref{fig_one_loop_electron_self_energy}, whereas the leading 
vertex-correction contribution starts at the two-loop order, as can be 
seen in fig.~\ref{fig_vertex_correction}.

  Therefore $\Delta_\mathrm{ex}$ in 
eqns.~(\ref{eq:electron_self_energy_dressed},%
\ref{eq:electron_self_energy_dressed_another}) cannot be interpreted as 
the bare vertex plus perturbative corrections of the form shown in 
fig.~\ref{fig_vertex_correction}.
  Equations~(\ref{eq:electron_self_energy_dressed},%
\ref{eq:electron_self_energy_dressed_another}) must instead be solved 
self-consistently.
  We shall do so using the Gribov equation formalism
\cite{gribovlargeshort,gribovquarkconfinement,gribovlectures,yurireview}.

 \section{Gribov equations}
 \label{sec_griboveqn}

  Following 
refs.~\cite{gribovlargeshort,gribovquarkconfinement,gribovlectures,yurireview}, 
we now apply
 \begin{equation}
  \frac{\partial^2}{\partial q_0^2}-
  \frac1{u^2}\frac{\partial^2}{\partial\mathbf{q}^2}
  \equiv \partial^2
 \end{equation}
  to eqn.~(\ref{eq:electron_self_energy_dressed}).

  Let us assume that the variations of $\Delta_\mathrm{ex}(k,q)$ with 
respect to $k-q$ are small, so that we need only consider the effect of 
applying $\partial^2$ on $D_\pm(q-k)$.
  This is correct in the weak coupling case, but becomes an 
approximation in the strong coupling case, as we find that 
$\Delta_\mathrm{ex}$ tends to fall at large energies as $1/E$.
  Even so, one expects that the structure of $G_\pm$ is determined 
mostly by soft-exchange contributions, $k-q\ll k$, in which case 
$\Delta_\mathrm{ex}(k,q)\approx\Delta_\mathrm{ex}(k)$ is a good 
approximation.
  Accepting this, the integral sign then disappears because of the 
following identity:
 \begin{equation}
  \partial^2
  \frac1{(q_0-k_0)^2-u^2(\mathbf{q}-\mathbf{k})^2+i0}=
  \frac{4\pi^2i}{u^3}\delta^{(4)}(q-k).
  \label{eqn_delta_emergence}
 \end{equation}
  We then obtain
 \begin{equation}
  \partial^2\Sigma_+(q)
  =\frac{(f^{-1}\Delta_\mathrm{ex}(q))^2}{4\pi^2\hbar^3u^3} G_-(q),
  \label{eqn_gribov_oneloop}
 \end{equation}
  and the same for $\Sigma_-$ when $G_-$ is replaced by $G_+$ on the 
right-hand side.

  Our approximation is that $\psi$ moves much more slowly than $\phi$ 
does. 
  We may therefore take the large $u$ limit of 
eqn.~(\ref{eqn_gribov_oneloop}), in which case only the energy 
derivative remains:
 \begin{equation}
  -\frac{\partial^2}{\partial q_0^2} G^{-1}_+(q)
  =\frac{(f^{-1}\Delta_\mathrm{ex}(q))^2}{4\pi^2\hbar^3u^3} G_-(q).
  \label{eqn_gribov_nr_step_a}
 \end{equation}
  We have replaced $\Sigma_+$ with $-G_+^{-1}$ on the right-hand side 
because the double energy derivative of the bare inverse propagator 
vanishes for Schr\"odinger fields.

  In order to simplify the equations, let us introduce a new variable 
$\omega_f=2\pi f(\hbar u)^{\nicefrac32}$, which has the dimension of 
energy.
  We obtain
 \begin{equation}
  \left\{
   \begin{array}{l}
   \vspace{1mm}
   (G^{-1}_+)^{\prime\prime}=-\omega_f^{-2}(G^{-1}_--G^{-1}_+)^2 G_-,\\
   (G^{-1}_-)^{\prime\prime}=-\omega_f^{-2}(G^{-1}_--G^{-1}_+)^2 G_+.
   \end{array}
   \right.
  \label{eqn_gribov_nr}
 \end{equation}
  Prime refers to the energy derivative $\partial/\partial q_0$.

 \section{Solution to Gribov equations}
 \label{sec_gribov_solution}

  Let us proceed to solve eqns.~(\ref{eqn_gribov_nr}).
  We first consider the expression
 \begin{equation}
  G_+(G_+^{-1})^{\prime\prime}-G_-(G_-^{-1})^{\prime\prime}=0,
 \end{equation}
  which follows from eqns.~(\ref{eqn_gribov_nr}).
  We then use the identity $x^{-1}x^{\prime\prime}\equiv (\ln 
x)^{\prime\prime}+((\ln x)^{\prime})^2$:
 \begin{equation}
  (\ln (G_+^{-1}/G_-^{-1}))^{\prime\prime}+
  ((\ln G_+^{-1})^\prime)^2-((\ln G_-^{-1})^\prime)^2=0.
 \end{equation}
  Let us define $z=G_+^{-1}/G_-^{-1}$:
 \begin{equation}
  (\ln z)^{\prime\prime}+
  (\ln z)^\prime(\ln (G_+^{-1}G_-^{-1}))^\prime=0.
 \end{equation}
  Integrating this expression once yields
 \begin{equation}
  (\ln z)^\prime=c_0G_+G_-.
 \end{equation}
  $c_0$ is a constant of integration with the dimension of energy.
  It is easy to see that $c_0=\Delta_0$ in the weak-coupling limit.
  Let us therefore denote it as such.
  We then obtain
 \begin{equation}
  G_-^2=\Delta_0^{-1}z^\prime, \quad G_+^2=\Delta_0^{-1}z^\prime/z^2.
  \label{eq:gribovConstofIntegration}
 \end{equation}

  Let us make use of eqn.~(\ref{eq:gribovConstofIntegration}) to 
eliminate the energy derivatives in eqns.~(\ref{eqn_gribov_nr}). This 
gives us, almost trivially,
 \begin{equation}
  G_-^3\frac{d^2}{dz^2}G_-=(\omega_f\Delta_0)^{-2}(z+z^{-1}-2).
  \label{eq:gribovGvsz}
 \end{equation}
  The equation for $G_+$ is obtained by substituting $G_+$ for $G_-$ and 
$z^{-1}$ for $z$.

  Let us proceed to solve eqn.~(\ref{eq:gribovGvsz}) numerically. 
  The definition of $z$ is such that the divergences of $G_\pm$ map to 
$z\to\infty$ and $z\to0$.
  We would like to replace it with some variable where both 
singularities appear at a finite value.

  In this regard, we notice that the left-hand side of 
eqn.~(\ref{eq:gribovGvsz}) has a conformal symmetry with respect to 
transformations of $z$. This symmetry becomes more manifest when we 
return to eqns.~(\ref{eqn_gribov_nr}) and now eliminate $G_-$ and $G_+$ 
using eqn.~(\ref{eq:gribovConstofIntegration}).
  After some elementary algebra, we obtain
 \begin{equation}
  \frac{z^{\prime\prime\prime}}{2z^\prime}-
  \frac34\left(\frac{z^{\prime\prime}}{z^\prime}\right)^2
  =\omega_f^{-2}(z+z^{-1}-2).
  \label{eq:gribovODE}
 \end{equation}
  We then see that the left-hand side of this equation is invariant 
under the M\"{o}bius transformation:
 \begin{equation}
  z\longrightarrow w=\frac{az+b}{cz+d}.
 \end{equation}

  Let us adopt
 \begin{equation}
  w=\frac{1+z}{1-z},\quad z=-\frac{1-w}{1+w}.
  \label{eq:mobius}
 \end{equation}
  The singularities now map to $w=\pm1$.
  We also define
 \begin{equation}
  y=
  \left(\frac{\omega_f\sqrt3}4\frac{dw}{dE}\right)^{\nicefrac12}
  =\left(\frac{\omega_f\Delta_0\sqrt3}{2\Delta_\mathrm{ex}^2}
  \right)^{\nicefrac12}.
  \label{eq:gribov_ydef}
 \end{equation}
  In terms of these variables, eqn.~(\ref{eq:gribovODE}) reduces to
 \begin{equation}
  y^3\frac{d^2y}{dw^2}=-\frac{3}{4(1-w^2)}.
  \label{eq:gribovyvsw}
 \end{equation}
  This may be derived also by applying eqn.~(\ref{eq:mobius}) directly 
to eqns.~(\ref{eq:gribovGvsz}) and (\ref{eq:gribovConstofIntegration}).

  By the symmetry between $G_\pm$, we have one boundary condition, 
namely $dy/dw=0$ at $w=0$. The other boundary condition is $y=y_0$ at 
$w=0$, where $y_0$ is a parameter. Since $y$ is proportional to 
$\sqrt{\Delta_0}/\Delta_\mathrm{ex}$, large $y_0$ corresponds to weak 
coupling and small $y_0$ to strong coupling.

\begin{figure}[ht]
  \centerline{\includegraphics[width=9cm]{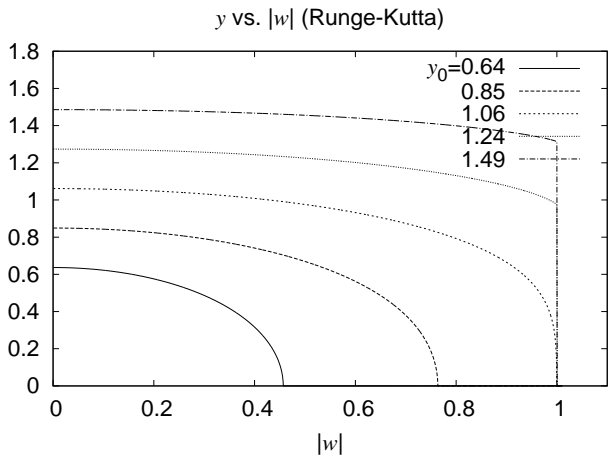}}
  \caption{Numerical results for $y$ versus $w$, for five representative 
values of $y_0=y(0)$.
  \label{fig_gribovyvsw}}
 \end{figure}

  We solved eqn.~(\ref{eq:gribovyvsw}) numerically using the above 
boundary conditions.
  The result is shown in fig.~\ref{fig_gribovyvsw}.
  The numbers were obtained using the classical Runge--Kutta method with 
the step size of 0.0005, and were found to be stable against 
modifications of the step size.

  There is a critical value $y_0^\mathrm{crit}$ of $y_0$, where $y$ 
vanishes when $w=\pm1$, with the limiting behaviour $y\to(1-w^2)^{1/4}$.
  The behaviour of the solution changes above and below 
$y_0^\mathrm{crit}$.
  We have found numerically that $y_0^\mathrm{crit}=1.061$ to three 
decimal places.

 \section{Analysis of the solutions}
 \label{sec_analysis}

  We would now like to discuss the nature of the solutions both below 
and above $y_0^\mathrm{crit}$.

\begin{figure}[ht]
  \centerline{\includegraphics[width=9cm]{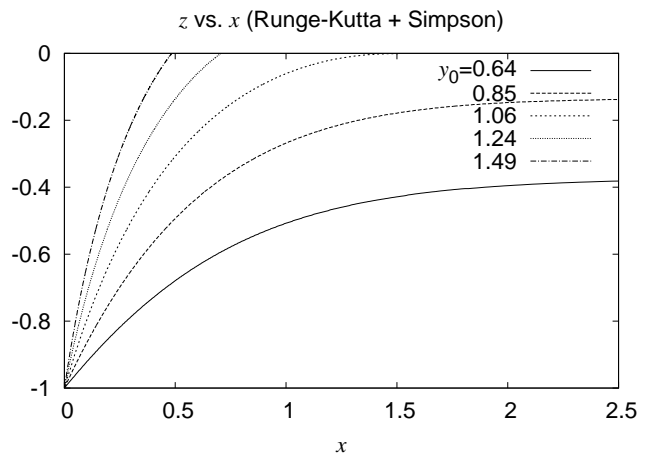}}
  \caption{Numerical results for $z$ versus $x$, for five representative 
values of $y_0=y(0)$.
  \label{fig_gribovzvsx}}
 \end{figure}

  Let us define the scaled energy $x$ as 
 \begin{equation}
  x=\frac4{\sqrt3}\frac{E-E_0}{\omega_f},
  \label{eqn_x_definition}
 \end{equation}
  with the boundary condition $w=0$ at $x=0$.
  By eqn.~(\ref{eq:gribov_ydef}), we obtain
 \begin{equation}
  \int dx=\int\frac{dw}{y^2(w)}.
  \label{eqn_x_as_integral_of_y_inverse_squared}
 \end{equation}
  We can use this equation to convert the $y$ versus $z$ results into 
relations involving scaled energy. As an example, we show $z$ as a 
function of $x$ in fig.~\ref{fig_gribovzvsx}, which is calculated 
numerically using the data of fig.~\ref{fig_gribovyvsw} and using 
Simpson's rule for integration.

  Let us now discuss the three regions of $y_0$.

  \subsection{Subcritical case ($y_0>y_0^\mathrm{crit}$)}

  When $y_0$ is large, eqn.~(\ref{eq:gribovyvsw}) is solved by the 
following approximate solution when $\left|w\right|$ is not large:
 \begin{equation}
  y\approx y_0-
  \frac3{8y_0^3}\left[(1+w)\ln(1+w)+(1-w)\ln(1-w)\right].
  \label{eqn_subcritical_approximation}
 \end{equation}
  When $\left|w\right|$ is large, the limiting behaviour is given by 
$y\to A-B\left|w\right|$, with $A\approx y_0$ and $B\propto y_0^{-3}$.
  If $y_0$ is large, $B$ is small. Since $y=\mathrm{const.}$ corresponds 
to constant $\Delta_\mathrm{ex}$, this behaviour corresponds to the weak 
coupling limit, as expected.
  The system is as described by eqn.~(\ref{eqn_yukawa_lagrangian_weak}).

\begin{figure}[ht]
  \centerline{\includegraphics[width=7cm]{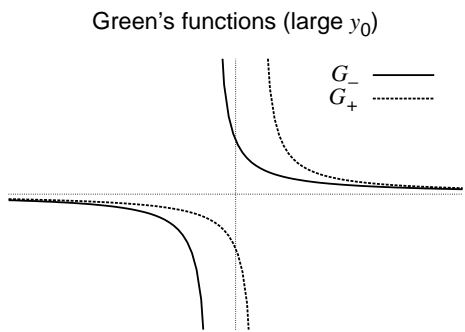}}
  \caption{The Green's functions in the weak-coupling limit as a 
function of scaled energy $x$.
  The normalization of axes is arbitrary.
  \label{fig_greens_approx_subcritical}}
 \end{figure}

  It can be seen that the singularity condition $w=\pm1$ is satisfied at 
finite and real $x$.
  Explicitly, $w=\pm1$ at near $x=\pm1/y_0^2$.
  The Green's functions have approximate poles on the real axis.
  Their behaviour is as shown in 
fig.~\ref{fig_greens_approx_subcritical}.
  The two Green's functions are separated by a constant exchange energy.

  Above the poles, $w$ grows linearly with $x$ at first, and saturates 
at $\left|w\right|=A/B$.
  This means, by the definition of $w$ and $z$, that the two Green's 
functions are renormalized asymmetrically. The asymmetry is proportional 
to $B/A$ and is therefore small when $y_0$ is large.

  \subsection{Supercritical case ($y_0<y_0^\mathrm{crit}$)}

  When $y_0$ is small, eqn.~(\ref{eq:gribovyvsw}) is solved by the 
following approximate solution when $w^2\ll1$.
 \begin{equation}
  y\approx\left(y_0^2-\frac{3w^2}{4y_0^2}\right)^{\nicefrac12}.
  \label{eqn_supercritical_approximation}
 \end{equation}
  This solution corresponds to the approximation $1-w^2\approx1$ in 
eqn.~(\ref{eq:gribovyvsw}).

  Equation~(\ref{eqn_x_as_integral_of_y_inverse_squared}) now yields
 \begin{equation}
  w=\frac{2}{\sqrt3}y_0^2\tanh\left(
  \frac{\sqrt3}{2}x\right).
  \label{eqn_supercritical_approximate_solution}
 \end{equation}
  So long as $y_0^2<\sqrt3/2$, the solution has no singularities 
$w=\pm1$ for real $x$.
  That is, $\psi$ is confined.

\begin{figure}[ht]
  \centerline{\includegraphics[width=7cm]{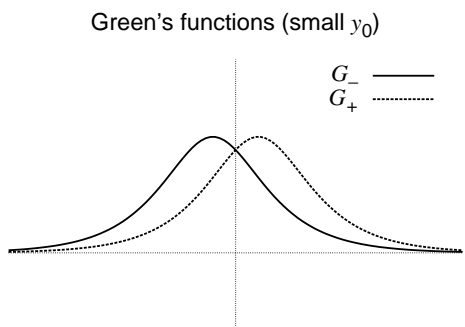}}
  \caption{The Green's functions in the strong-coupling limit as a 
function of $x$.
  The normalization of axes is arbitrary.
  \label{fig_greens_approx_supercritical}}
 \end{figure}

  Let us calculate $G_\pm$. 
Equation~(\ref{eqn_supercritical_approximate_solution}) implies
 \begin{equation}
  z=-\frac{\cosh\frac{\sqrt3}2(x+\delta_0)}{
  \cosh\frac{\sqrt3}2(x-\delta_0)},
 \end{equation}
  where $\delta_0=(2/\sqrt3)\tanh^{-1}(2y_0^2/\sqrt3)\approx4y_0^2/3$.
  The Green's functions are then given by
 \begin{equation}
  G^{-1}_\pm=\pm\cosh\frac{\sqrt3}2\left(x\pm\delta_0\right),
  \label{eqn_supercritical_approximate_greens}
 \end{equation}
  up to an irrelevant normalization.
  This result is shown in fig.~\ref{fig_greens_approx_supercritical}.
  The Green's functions decay rapidly away from $x\approx0$. The $\psi$ 
field only exists in a narrow region of energy.

  Singularities now appear off the real axis.
  The form of eqn.~(\ref{eqn_supercritical_approximate_greens}) implies 
a cyclic series of singularities, of which the ones nearest to the real 
axis are at
 \begin{equation}
  x=\delta_0\pm\frac{i\pi}{\sqrt3},\quad
  -\delta_0\pm\frac{i\pi}{\sqrt3}.
 \end{equation}
  Hence the positions of the singularities, and the decay rate of the 
false vacuum \cite{gribovquarkconfinement}, are determined primarily by 
$\omega_f$ by eqn.~(\ref{eqn_x_definition}).
  Note that the behaviour of the Green's functions very close to the 
singularities are modified because the approximation $w^2\ll1$ fails.

  Concerning the behaviour near $w=\pm1$, roughly the same 
considerations as the weak-coupling case apply, and the singularities in 
the strong coupling limit are asymptotically of the form of simple 
poles.
  One difference is in that, as can be seen from 
eqn.~(\ref{eqn_supercritical_approximation}), singularities $w=\pm1$ 
occur for pure imaginary $y$.
  This implies that, since $y^2=dw/dx$, the sign of $z$, that is, the 
relative sign between $G_\pm$, is reversed.

  Our findings are consistent with the discussion of 
ref.~\cite{gribovquarkconfinement}, in that in a confining theory, no 
real singularities appear, and complex singularities represent the decay 
of the false vacuum into a vacuum with vacant negative energy states and 
occupied positive energy states.
  These states emerge because $\psi_+\psi_-$ pairs are bound together 
into supercritical bound states.

  However, the appearance of the cyclic series of singularities requires 
explanation. This is due to the cyclic series of states that occur in 
the critical case, to be discussed in the following.

  \subsection{Critical case ($y_0=y_0^\mathrm{crit}$)}

  Last of all, let us discuss the critical case $y_0=y_0^\mathrm{crit}$.

\begin{figure}[ht]
  \centerline{\includegraphics[width=7cm]{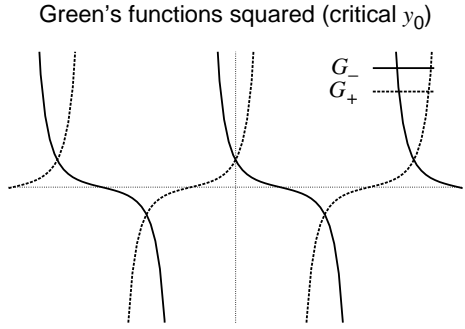}}
  \caption{Approximate Green's functions squared in the critical case as 
a function of $x$.
  The normalization of axes is arbitrary.
  \label{fig_greens_approx_critical}}
 \end{figure}

  The approximate solution is now given by $y=(1-w^2)^{\nicefrac14}$.
  This leads to
 \begin{equation}
  w=\sin x,\quad
  z=\frac{\sin x-1}{\sin x+1}=-\tan^2\left(\frac{x-\nicefrac\pi2}2\right).
  \label{eqn_gribov_nr_approx_solution}
 \end{equation}
  This implies that, by eqn.~(\ref{eq:gribovConstofIntegration}),
 \begin{eqnarray}
  G_-&=&(\cos x)^{\nicefrac12}(\sin x+1)^{-1},
  \label{eqn_gribov_nr_approx_solution_g_a}\\
  G_+&=&(\cos x)^{\nicefrac12}(\sin x-1)^{-1},
  \label{eqn_gribov_nr_approx_solution_g_b}
 \end{eqnarray}
  up to an irrelevant overall normalization.
  This behaviour is shown in fig.~\ref{fig_greens_approx_critical}.
  Note that we have plotted Green's functions squared here unlike in the 
previous two figures.

  We see that the solution is cyclic.
  Near the singularities, e.g., $x=\pi/2-\epsilon$, these functions 
behave as
 \begin{equation}
  G_-\to\frac{\sqrt{\epsilon}}2,\quad 
  G_+\to-2\epsilon^{-\nicefrac32}.
  \label{eq:x_to_zero_behaviour}
 \end{equation}
  This is clearly a rather exotic behaviour.

  Let us recall the Levinson theorem \cite{levinson,levinson2} which 
states that the number of states in between energies $E_\mathrm{A}$ and 
$E_\mathrm{B}$ is given by the difference in the phase shifts $\delta$:
 \begin{equation}
  N=\frac1\pi\left[\delta\left(E_\mathrm{B}\right)-
  \delta\left(E_\mathrm{A}\right)\right].
 \end{equation}
  In this sense, we can say that there are $-\nicefrac12$ $\psi_-$ 
states and $+\nicefrac32$ $\psi_+$ states at $x=\nicefrac\pi2$.
  Similarly, there are $-\nicefrac12$ $\psi_+$ states and $+\nicefrac32$ 
$\psi_-$ states at $x=\nicefrac{3\pi}2$.
  This is if we adopt the convention of moving singularities above the 
real axis for $E>\mu_\mathrm{F}$ and below the real axis for 
$E<\mu_\mathrm{F}$.
  But this convention cannot be right when it produces negative number 
of states.

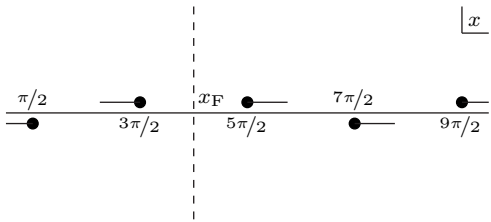
\begin{figure}[ht]
  \centerline{
  \begin{picture}(200,100)
   \Line(10,50)(190,50)
   \Line(180,80)(180,90)
   \Line(180,80)(190,80)
   \Text(183,83)[bl]{$x$}
   \DashLine(80,10)(80,90){3}
   \Text(20,55)[c]{$\nicefrac\pi2$}
   \GCirc(20,46){2}{0}  \Line(20,46)(10,46)
   \Text(60,45)[c]{$\nicefrac{3\pi}2$}
   \GCirc(60,54){2}{0}  \Line(60,54)(45,54)
   \Text(82,55)[l]{$x_\mathrm{F}$}
   \Text(100,45)[c]{$\nicefrac{5\pi}2$}
   \GCirc(100,54){2}{0} \Line(100,54)(115,54)
   \Text(140,55)[c]{$\nicefrac{7\pi}2$}
   \GCirc(140,46){2}{0} \Line(140,46)(155,46)
   \Text(180,45)[c]{$\nicefrac{9\pi}2$}
   \GCirc(180,54){2}{0} \Line(180,54)(190,54)
  \end{picture}
  }
  \caption{\label{fig_analytic}
  Moving away real-axis singularities, for $G_-$ in the critical case. 
$x_\mathrm{F}$ corresponds to the Fermi energy.}
 \end{figure}

  The negative-number states need to be occupied or vacated, and become 
occupied positive-energy states and vacant negative-energy states, and 
therefore their cuts are located on the other side of the real axis, as 
shown in fig.~\ref{fig_analytic}.

  Let us now say that we excite a $\psi_+$ state at $x=0$.
  This $\psi_+$ can emit a critical $\psi_+\psi_-$ pair without any cost 
of energy, and change into $\psi_-^\dagger$ at $x=0$. Thus there is a 
mixing between $\psi_+$ and $\psi_-^\dagger$, and therefore the presence 
of a $\psi_-$ hole (i.e., occupied positive energy state) at the same 
energy as $\psi_+$ is explained.

  Our convention has been that $\psi_+$ states are at a higher energy 
than $\psi_-$ states. In this case, the energy difference between the 
states in terms of $x$ is $\pi$. As a result of having both $\psi_+$ and 
$\psi_-$ states at the same energy, it follows that there are states 
both $\pi$ above and $\pi$ below $x=0$. The same applies to states at 
$x=\pi$ and so, continuing ad infinitum, we come to the conclusion that 
there must indeed be a tower of states, though this seems strange and 
exotic.

 \section{Discussion}
 \label{sec_discussion}

 \subsection{Nature of the critical pairs}

  From the structure of the self-energy diagram, and of the solutions, 
the complex singularities can be seen to correspond to $\psi_+\psi_-$ 
(super-)critical bound states.
  These pairs are strange objects: they are not Cooper pairs which are 
formed out of electrons that are near the Fermi surface.

  At least mathematically, the critical solution to Gribov equations 
implies the following:
 \begin{enumerate}
  \item A real $\psi_+$ state is accompanied by a $\psi_-^\dagger$ hole 
state, because of the emergence of $\psi_+\psi_-$ critical states.
  \item $\psi_+\psi_-$ states are formed by pairing a $\psi_+$ state 
with the excitation $(\psi_-^\dagger)^\dagger$ of the accompanying 
$\psi_-^\dagger$ state.
 \end{enumerate}
  This is an unpleasant chicken-or-egg situation. A more intuitive 
statement is that pairs are formed by the interaction of, let us say, a 
real $\psi_+$ with a virtual $\psi_-$.

  This should be classified as a spin polaron state \cite{spinpolaron}.
  However, unlike the string polaron \cite{stringpolaron} where 
confinement is due to interaction which grows at long distances, we have 
a dynamics which is described by the temporal decay of the false vacuum, 
i.e., soft confinement.

  Physically, when fermionic spin is excited, it will decay by emitting 
a magnon, and will not behave like an itinerant spin.

  Whether or not the condensation of supercritical pairs leads to 
superconductivity depends on whether these pairs are delocalized (and 
hence superconducting) or localized (and hence insulating).
  The equations suggest delocalized pairs, but presumably there are both 
possibilities, depending on the separation between charge carriers, for 
instance, which information is omitted in our framework.
  Since the supercritical pairs are not Cooper pairs, the 
superconductivity, even if it is realized, would be of a very different 
type to conventional ones.

 \subsection{Comparison with real spin systems}

  One important issue concerns whether our system, which is equipped 
with an $\omega\propto\left|\vec k\right|$ spin-wave coupled to slow 
fermions, describes some real spin system.

  On the outset, this appears unlikely, because a linear dispersion 
relation usually occurs in antiferromagnetic systems whereas the Yukawa 
coupling is appropriate to ferromagnetic systems.

  One possibility concerns the case of doped antiferromagnetic 
insulators.
  In this case, the carriers will tend to move by hopping and so their 
velocity will typically be much slower than the spin-wave velocity.
  Furthermore, if the system has a sufficiently metallic character, one 
may expect intuitively that the Yukawa interaction becomes a reasonable 
description.

  This suggests that somewhere in the metal--insulator transition phase 
diagram, there may exist a region in which the $\psi_+\psi_-$ condensate 
arises, corresponding to exotic insulator or superconductor.

  This has not been observed in three spatial dimensions where, to our 
knowledge, there are no reported instances of antiferromagnetic metal to 
antiferromagnetic insulator transition to start with\footnote{We thank 
H.~Eisaki and I.~Hase for valuable discussions on this point.}.

  On the other hand, in two spatial dimensions, where our results are 
not directly applicable, high-$T_c$ superconductivity does occur in 
cuprate \cite{highTc,highTc_book} and iron pnictide \cite{FeAs} systems.
  Further work is desirable.

 \subsection{Comparison with QCD}

  Despite close similarities, there are some differences in the findings 
of our study as compared to the QCD counterpart 
\cite{gribovlargeshort,gribovquarkconfinement}.

  In the QCD case, the gluon is responsible for binding together $q\bar 
q$ pairs supercritically, and it is the pion exchange that crucially 
modifies the analyticity of the quark Green's functions.

  Our system differs in that it is magnon exchange that binds together 
$\psi_+\psi_-$ pairs supercritically, and the magnon exchange by itself 
already modifies the analyticity of the $\psi$ Green's functions.
  The main difference is that unlike the QCD case where the pion 
interaction is at least in principle determined self-consistently, 
$f=2v_h$ is a free parameter here.
  Our study is analogous to solving the quark Gribov equation without 
the gluon exchange contribution, assuming that chiral symmetry breaking 
has nevertheless occurred.

  The tower of states that occurs in our case at critical coupling 
apparently has no counterpart in QCD, even though it is suggestive of 
meson trajectories.

 \section{Conclusion}

  We have written down a coupled system of Gribov equations for slow 
fermions $\psi$ interacting with fast $\omega\propto\left|\vec k\right|$ 
magnons $\phi$ in three spatial dimensions.

  The solution exhibits qualitatively different behaviour depending on 
the strength of interaction.
  When the interaction is sufficiently strong, supercritical pairs of 
$\psi$ condense, much like the corresponding situation in Gribov's 
light-quark confinement scenario 
\cite{gribovlargeshort,gribovquarkconfinement,gribovlectures,yurireview}.
  This corresponds to a new spin polaron state and may lead to exotic 
superconductivity.

  A particularly exotic behaviour is the presence of an infinite tower 
of (real or complex) states.
  This behaviour arises because a state at a certain energy level is 
necessarily accompanied by a spin-flipped state either above or below 
it.
  The unusual occupation of states then requires the presence of this 
unusual tower of states.

 \vspace*{1cm}


 \begin{acknowledgements}

  We thank H.~Asai, H.~Eisaki, I.~Hase, M.~Hashimoto and S.~Kawabata for 
stimulating discussions.

 \end{acknowledgements}

 \end{document}